\documentclass[twocolumn,twoside,slac_two]{revtex4}
\usepackage{graphicx}
\usepackage{fancyhdr}
\usepackage{hyperref}
\hypersetup{
    colorlinks=true,
    urlcolor=magenta,
}
\usepackage{amsmath}

\begin{document}

\title{Search for Cosmic Ray Anisotropy with the Alpha Magnetic
  Spectrometer on the International Space Station}

%

\author{G. La Vacca and the AMS-02 Collaboration} 
\affiliation{INFN Milano-Bicocca, Piazza della Scienza 3, 20125
  Milano, Italy}
%

\begin{abstract}
The search for cosmic ray anisotropy has been performed using
particles collected by the Alpha Magnetic Spectrometer after the first
five years of operation. The positron to electron ratio is consistent
with isotropy at all energies and angular scales. The limit on the
dipole anisotropy at the 95\% confidence level for energies above 16
GeV is $\delta<0.020$ for positrons and $\delta<0.006$ for
electrons. In the case of high rigidity protons the upper limit in the
rigidity range from 80 to 1800\,GV is $\delta<0.003$. No indication of
seasonal excess is observed for all particle species within the
present statistics.
\end{abstract}

\maketitle

\thispagestyle{fancy}

Recent precise measurements of Cosmic Ray (CR) fluxes performed by the
Alpha Magnetic Spectrometer (AMS-02) revealed many structures in the
CR spectra that are fundamental for the understanding of the CR origin
problem. In particular the proton spectral shape showed a hardening of
the spectral index at rigidities above ~100\,GV
\cite{AMSproton}. Besides the slope of the positron fraction decreases
logarithmically with energy above 30\,GeV and above ∼200\,GeV the
positron fraction is no longer increasing with energy
\cite{AMSposfract}. This behaviour originates from the electron and
positron fluxes dependence on energy. Above ∼20\,GeV and up to
200\,GeV the electron flux decreases more rapidly with energy than the
positron flux, that is, the electron flux is softer than the positron
flux \cite{AMSposele}.

All these features cannot be fully explained within the current
physical knowledge. However they may be connected to new phenomena
which, in principle, could induce some degree of anisotropy in the CR
arrival directions. Therefore, besides CR spectra, it is worth to
characterize the properties of the observed angular distribution of
the galactic CR in order to provide complementary information for
constraining both the origin and the propagation mechanism of CRs in
the heliosphere and in the Galaxy.

In this work we analized the sample of the first five years of the
AMS-02 data, searching for relative anisotropies in the galactic CR
incoming directions. The analysis has been performed both at the
detector altitude and after reconstructing the CR trajectories in the
geomagnetic field. A systematic study of the seasonal dependence of
the anisotropy has been performed.

\section{The AMS-02 Detector}
AMS-02 is a TeV multipurpose CR detector, designed to conduct a long
term mission (about 14 years). It was installed at nearly 400\,km
altitude on board of the International Space Station (ISS) on 19 May
2011 and since then it is operating 24/7, so far collecting more than
40 billion events of galactic CR.

The detector \cite{Ting201312} consists of nine planes of precision
silicon tracker with two outer planes, 1 and 9, and the inner tracker,
planes 2 to 8; a transition radiation detector (TRD); four planes of
time of flight (TOF) counters; a permanent magnet; an array of
anticoincidence counters (ACC), inside the magnet bore; a ring imaging
\v{C}erenkov detector (RICH); and an electromagnetic calorimeter
(ECAL).

\section{Data Selection}
The quality criteria for the electron and positron selection for this
analysis follows that described in \cite{AMSposele} and in
\cite{AMSposfract}.
The reduction of the proton background in the identification of the
positron and electron samples is achieved by means of the TRD, the
ECAL and the tracker. Events are selected by requiring a track in the
TRD and in the tracker, a cluster of hits in the ECAL, and a measured
velocity $\beta\sim 1$ in the TOF consistent with a downward-going
Z$=1$ particle. Protons are rejected by requiring a good
energy-momentum matching and explicit cuts on the ECAL and TRD
estimators. 
The remaining sample contains 70000 primary positrons, 920000
electrons and a negligible amount of protons.

The selected events are grouped into 5 cumulative energy ranges from
16 to 350\,GeV according to their measured energy in the ECAL. The
minimum energies are 16, 25, 40, 65 and 100\,GeV, respectively.

Proton selection follows the one described in \cite{AMSproton}. The
first step of the selection requires the preselection of events in
which the velocity was measured by at least three TOF layers being
consistent with down-going particles, and the linearly extrapolated
trajectory of the TOF hit positions passing both tracker layer 1 and
9. Then further cuts are made requiring the charge consistency with
Z$=1$ particle and requiring the track to pass through layer 1 and 9
and to satisfy additional track fitting quality criteria.

In addition, to select only primary CRs, well above the geomagnetic
cutoff, the measured rigidity is required to be greater than 1.2 times
the maximum geomagnetic cutoff within the AMS field of view.

\section{Methodology and sky maps}
\begin{figure*}
  \includegraphics[width=80mm]{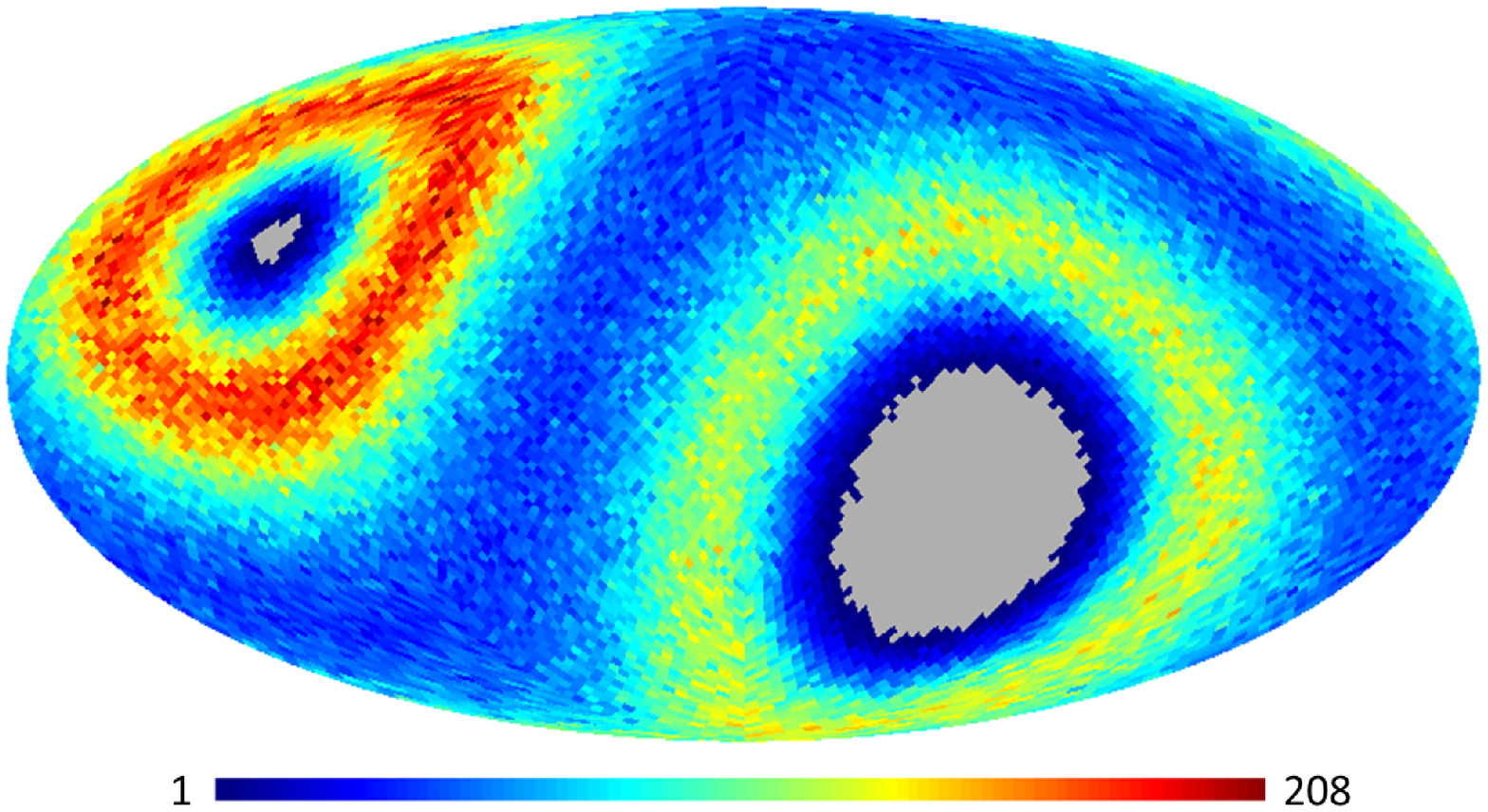}
  \includegraphics[width=80mm]{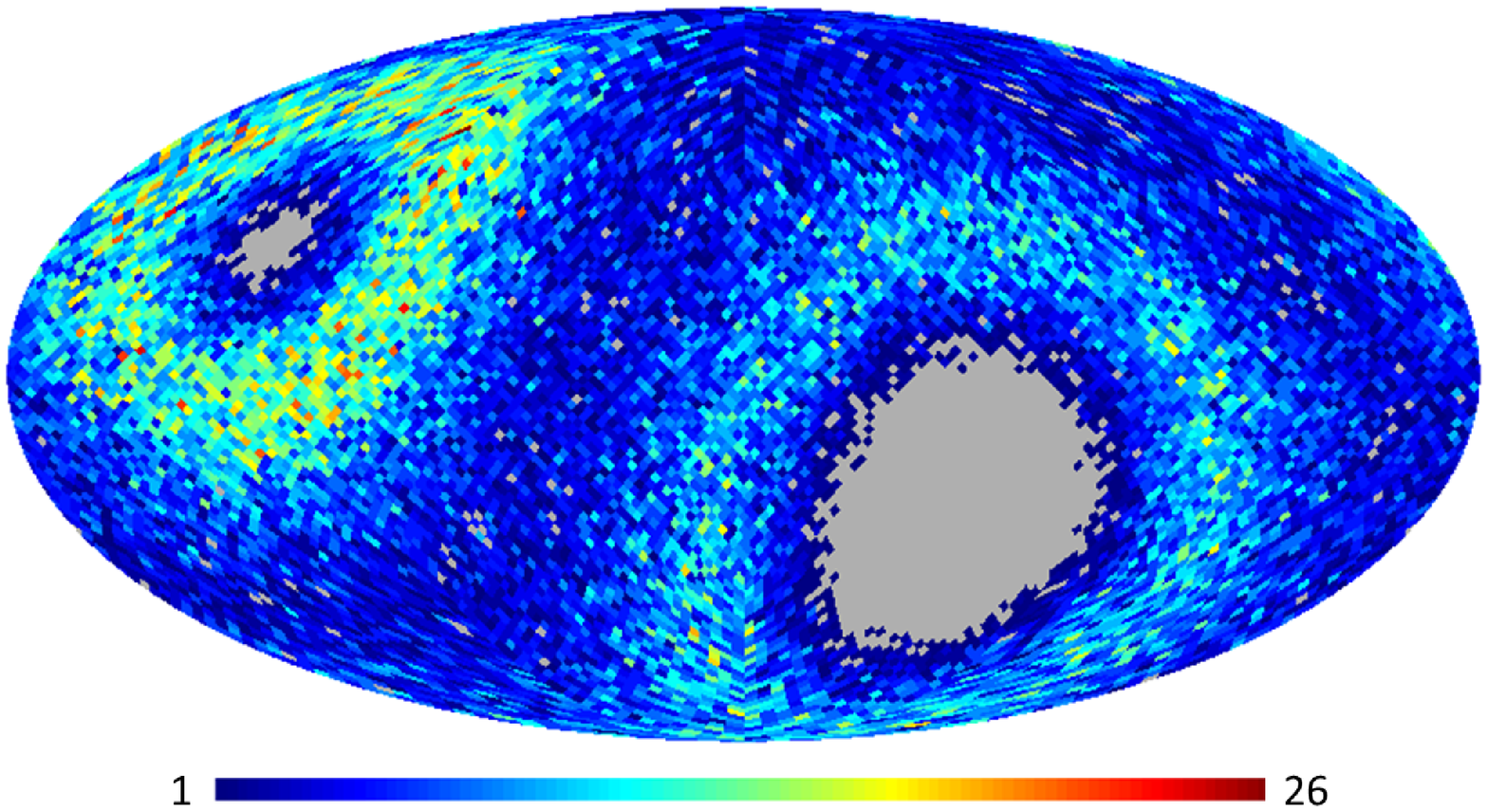}
  \caption{Sky maps showing the arrival directions of selected
    electrons (left) and positrons (right) in the energy bin
    [16,350]\,GV in galactic coordinates as observed at AMS
    altitude. The color code reflects the number of events per bin.}
  \label{maps}
\end{figure*}

The CR arrival directions are used to build sky maps in galactic
coordinates $(l,b)$ containing the number of observed particles. The
sky maps are built using the {\sc
  healp}ix \footnote{http://healpix.sourceforge.net} pixelization,
which guarantees a regular sampling of the CR angular
distribution. The maps corresponding to electrons and positrons in the
energy range from 16 to 350 GeV are displayed in Fig.~\ref{maps}.
From this figure it is clear that AMS-02 is able to perform a nearly
full-sky observation and therefore a three-dimensional measurement of
the tiny CR anisotropy signals. Moreover, due to the ISS orbit
inclination of 51.6$^\circ$ relative to the Earth equator, combined
with the AMS-02 attitude on the ISS, the sky coverage of the AMS-02
exposure is non-uniform, being larger in the North Geographic Pole
region than in the South. All CR species share the same exposure
property.

The analysis of anisotropies requires the creation of reference maps
that represent the null-hypothesis or the isotropic sky as seen by the
detector. In this work we present results about the study of the {\it
  relative} anisotropy in which other CR species are used as reference
for the sample of interest, such as electrons or protons for
positrons. Another option discussed here is the use of a CR
distribution as reference for the same species but at a different
rigidity bin, such as low rigidity protons as reference for high
rigidity protons.

Using the distribution of events in the reference maps, a likelihood
fit procedure has been set up to compare the species under study to
the reference sky map. The fit also takes into account the differences
in the exposure for different rigidities due to the different
geomagnetic cutoff. This procedure proved to be stable against
different map resolutions and sample statistics. The flux is expanded
in spherical harmonics
\begin{equation}
\frac{\phi_i - \langle\phi_i\rangle}{ \langle\phi_i\rangle} =
\sum^{\ell_{max}}_{\ell=0} \sum^{\ell}_{m=-\ell} a_{\ell m} Y_{\ell
  m}(l_i,b_i),
\label{expan}
\end{equation}
where $(l_i,b_i)$ is the position of the $i^{th}$ pixel in galactic
coordinates. The dipole for $\ell=1$ is fully described by three
orthogonal functions aligned in galactic coordinates with interesting
directions: $Y_{10}$ along the North-South (NS) direction
perpendicular to the galactic plane, $Y_{11}$ along the
Forward-Backward (FB) direction with respect to the galactic center
and $Y_{1-1}$ along the East-West (EW) direction tangent to the orbit
of the sun around the galactic center. It is then worth to study these
three directions both separately and combined to determine the total
omnidirectional dipole magnitude
\begin{equation}
\delta=\sqrt{\rho_{NS}^2+\rho_{FB}^2+\rho_{EW}^2}. 
\label{delta}
\end{equation}
where the three dipole coefficients are defined as
\begin{equation}
  \rho_{NS}=\sqrt{\frac{3}{4\pi}} a_{10} ~
  \rho_{FB}=\sqrt{\frac{3}{4\pi}} a_{11} ~
  \rho_{EW}=\sqrt{\frac{3}{4\pi}} a_{1-1} ~.
\end{equation}

\section{Anisotropy in positron to electron ratio}
\label{PosEle}  
\begin{figure}[t]
  \includegraphics[width=65mm]{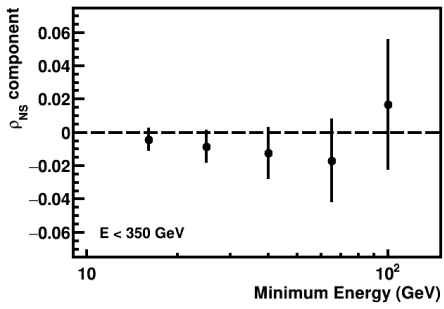}
  \caption{Amplitudes of the $\rho_{NS}$ dipole component obtained for
    different energy ranges in the positron over electron case.}
  \label{PosEle_RhoNS}
\end{figure}

\begin{figure}[t]
  \includegraphics[width=65mm]{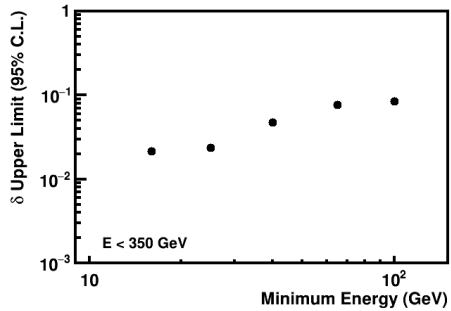}
  \caption{Upper limits at 95\% C.L. of the omnidirectional dipole
    intensity in the positron to electron ratio.}
  \label{PosEle_Delta}
\end{figure}

In Fig.~\ref{PosEle_RhoNS} the results corresponding to a dipole
contribution $\rho_{NS}$ in the case of positron to electron ratio are
shown as function of the minimum of the energy bin. The points at
every energy show no significant deviation from isotropy. Similar
plots can be obtained on the amplitudes of the other dipole
components.
The 95\% C.L. upper limits on the dipole amplitude for different
energies are reported in Fig.~\ref{PosEle_Delta}. The limit obtained
for the energy range from 16 to 350\,GeV is $\delta<0.020$. This limit
is consistent with the one reported in \cite{AMSposfractFirst}, the
only improvement due to the increased statistics after five years.

Further checks have been performed on the variation of the upper
limits as function of seasons (three months time binning), looking for
possible time dependence of the signal. The full data-taking period is
divided into 16 seasons and the analysis is repeated on the individual
samples. Even in this analysis it can be shown that no sigificant
deviation from isotropy is found.

\section{Anisotropy at magnetosphere border}
\label{BackTr}  
Since AMS is located on board of the ISS in a Low Earth Orbit (LEO),
the charged CR detected by AMS before interacting with the atmosphere
are exposed to the influence of the magnetosphere which deviates their
trajectories denpending on their rigidities and charges. This could in
principle have an effect on the measurement of the anisotropy.

In order to estimate the sensitivity to a dipole contribution due to
geomagnetic effects we compared the results of the analysis at ISS
altitude with the ones obtained using the asymptotic CR directions at
the entrance of the magnetosphere after backtracing. We backtraced CRs
from the top of AMS out to 25 Earth’s radii using the most recent
International Geomagnetic Reference Field (IGRF) model \cite{IGRF12}
with external nonsymmetric magnetic fields
\cite{JGRA:JGRA18039,JGRA:JGRA17702}, which describes the Earth
magnetosphere during both quiet and active solar periods.

Since the deviation introduced by the magnetosphere with respect to
the asymptotic direction depends on the CR charge, sky maps from
opposite charge particles cannot be compared at magnetosphere
border. The analysis described in Sec.~\ref{PosEle} has thus been
repeated using protons as reference for positrons, while the reference
map for electrons was the proton sky map obtained after backtracing
protons with negative charge.

A pure proton sample covering the same angular acceptance as the
positron and electron samples is achieved by reversing the cuts on the
ECAL and TRD estimators. The selected protons are classified into the
energy ranges defined in Sec.~\ref{PosEle} according to their measured
rigidity.

\begin{figure}[t]
  \includegraphics[width=65mm]{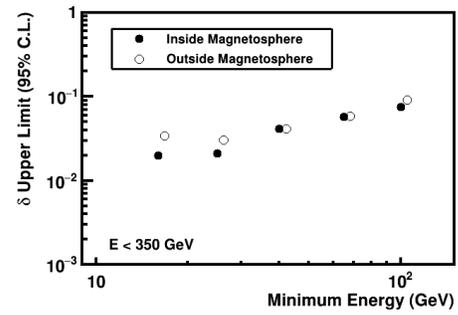}
  \caption{Upper limits at 95\% C.L. of the omnidirectional dipole
    intensity in the positron to proton ratio.}
  \label{PosPr_Delta}
\end{figure}

\begin{figure}[t]
  \includegraphics[width=65mm]{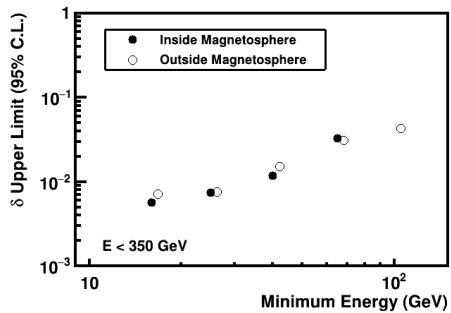}
  \caption{Upper limits at 95\% C.L. of the omnidirectional dipole
    intensity in the electron to inverted charge proton ratio.}
  \label{ElePr_Delta}
\end{figure}

As reported in Fig.~\ref{PosPr_Delta} the sensitivity to a dipole
anisotropy using the positron to proton ratio is compatible with that
obtained on the positron to electron analysis and the results are
consistent with those presented in Sec.~\ref{PosEle}. The comparison
between limits at ISS altitude and at magnetosphere border shows
compatibility and no significant deviation from isotropy.
Similar comments can be expressed for the results on the electron to
inverted charge proton ratio reported in Fig.~\ref{ElePr_Delta}. The
95\% C.L. upper limits on the dipole amplitude obtained for the energy
range from 16 to 350\,GeV is $\delta<0.006$ and is completely
consistent with isotropy and with the larger electron statistics with
respect to positrons.
%
The check for seasonal dependences in the results has been perfomed
following the same procedure as described in Sec.~\ref{PosEle}. No
seasonal variation are observed within the current statistics for both
the ratios.

\section{Anisotropy in high energy protons}
\label{HiRigPro}
The break of the proton spectra above 100GV could raise the question
whether this effect can be associated to a directional dependence of
the incoming energetic protons. In order to look for this effect the
sky map of protons in the range [40,80]\,GV, well above geomagnetic
cutoff, has been used as reference map for higher rigity protons.  For
this analysis protons are grouped into 5 cumulative rigidity bins from
80 to 1800\,GV according to their measured rigidity. The minimum
rigidities are 80, 150, 300, 500, 1000\,GV, respectively.

\begin{figure}[t]
  \includegraphics[width=65mm]{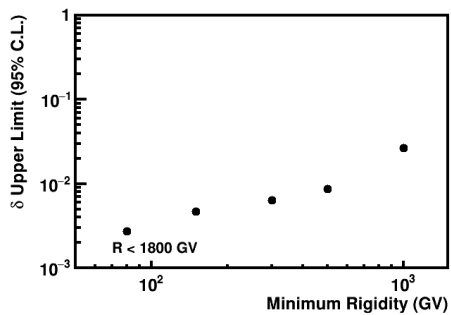}
  \caption{Upper limits at 95\% C.L. of the omnidirectional dipole
    intensity in the high rigidity proton to low rigidity proton
    ratio.}
  \label{PrPr_Delta}
\end{figure}

The 95\% C.L. upper limits on the dipole amplitude for different
minimium rigidities are reported in Fig.~\ref{PrPr_Delta}. The points
at every rigidity show no significant deviation from isotropy. The
limit obtained for the rigidity range from 80 to 1800\,GV is
$\delta<0.003$. Also in this case no seasonal effects are observed.

\section{Conclusions}
We presented the update of the results on CR anisotropy search with
AMS-02 detector after five years of data taking. Within the present
statistics no significant deviation from isotropy has been found in
all CR species considered, both at ISS altitude and at the
magnetosphere entrance. The 95\% confidence level limit obtained for
the energy range from 16 to 350 GeV is $\delta<0.020$ for the positron
sky map and $\delta<0.006$ for electrons. In the case of protons the
upper limit in the rigidity range from 80 to 1800\,GV is
$\delta<0.003$. No indication of seasonal excess is observed on the CR
anisotropy.

\bigskip 
\begin{acknowledgments}
This work has been supported by persons and institutions acknowledged
in \cite{AMSproton}. The authors acknowledge support by the state of
Baden-W\"urttemberg through bwHPC, and the German Reserarch Federation
(DFG) through grant no INST 39/963-1 FUGG. Some of the results in this
paper have been derived using the HEALPix (K.M. G\'orski et al., 2005,
ApJ, 622, p759) package.
\end{acknowledgments}

\bigskip 
\bibliography{ECRS_proc_template_LaTeX}

\end{document}